\documentclass[12pt]{article}
\usepackage{amsfonts,amssymb,amsmath}
\usepackage[dvips]{epsfig}
\begin{document}
\title{{\em{General Displaced $SU (1,1)$ number states - revisited}}}
\author{A. Dehghani\thanks{Email: alireza.dehghani@gmail.com, a\_dehghani@tabrizu.ac.ir}\\
{\small {\em Physics Department, Payame Noor University, PO Box
19395-3697 Tehran, Iran \,}} } \maketitle
\begin{abstract}
The most general displaced number states, based on the bosonic and
an irreducible representation( IREP) of the Lie algebra symmetry of
$su(1,1)$ and associated to the Calogero-Sutherland model are
introduced. Here, we utilize the Barut-Girardello displacement
operator instead of the Klauder- Perelomov counterpart, to construct
new kind of the displaced number states which can be classified in
nonlinear coherent states regime, too, with special nonlinearity
functions. They depend on two parameters, and can be converted into
the well known Barut-Girardello coherent and number states
respectively, depending on which of the parameters equal to zero. A
discussion of the statistical properties of these states is
included. Significant are their squeezing properties and anti
bunching effects which can be raised by increasing the energy
quantum number. Depending on the particular choice of the parameters
of the above scenario, we are able to determine the status of
compliance with flexible statistics. Major parts of the issue is
spent on something that these states, in fact, should be considered
as new kind of photon-added coherent states, too. Which can be
reproduced through an iterated action of a creation operator on new
{\em nonlinear Barut-Girardello coherent states}. Where the latter
carry, also, outstanding statistical features.
\\
\\
{\bf keywords:}  Photon-added coherent state, Sub-Poissonian,
Squeezing, Anti-bunching, Entanglement.

\end{abstract}

\section{Introduction}
In recent years, many authors have investigated new quantum states
of the electromagnetic field such as coherent states [1- 12] which
provide us with a link between quantum and classical mechanics and
nowadays pervade many branches of physics including quantum
electrodynamics, solid-state physics, and nuclear and atomic
physics, from both theoretical and experimental viewpoints. In
addition to CSs, squeezed states (SSs) are becoming increasingly
important. These are the non-classical states of the electromagnetic
field in which certain observables exhibit fluctuations less than in
the vacuum state \cite{Stoler}. These states are important because
they can achieve lower quantum noise than the zero-point
fluctuations of the vacuum or coherent states. Over the last four
decades there have been several experimental demonstrations of
nonclassical effects, such as the photon anti-bunching
\cite{Kimble}, sub-Poissonian statistics \cite{Short}, and squeezing
\cite{Matos1}.

Besides the above developments, on the one hand, special attention
has also been carried out to investigate the properties of the
displaced number states (DNSs),
\begin{eqnarray}
&&\hspace{-15mm}DNSs:\equiv\mathcal{D}_{KP}({z})|m,\lambda\rangle\nonumber\\
&&\hspace{-15mm}\mathcal{D}_{KP}(z):\equiv Klauder- Perelomov\hspace{2mm}type\hspace{2mm}of\hspace{2mm}Displacement\hspace{2mm}Operator\nonumber\\
&&\hspace{-15mm}|m,\lambda\rangle=
Arbitrary\hspace{2mm}Fock\hspace{2mm}States\nonumber\end{eqnarray}which
were given by Plebanski in 1954 \cite{Plebanski} and more analyzed
by Boiteux in 1973 \cite{Boiteux} under the general name semi
coherent states overall differences than what was presented
previously in \cite{Dodonov1} (the reader can find comprehensive
information about DNSs in \cite{Oliveria, Oliveria1}). It has been
shown that such states have interesting and unusual physical
properties. Since DNSs are obtained from a number states by adding a
non-zero value to field amplitude, the state becomes phase dependent
because of the phase of the displacement. The fact that these phase
dependent makes it interesting to study their phase properties
\cite{Tanas}. Also, the construction of DNSs is studied
\cite{dehghanif} corresponding with a free particle moving on
sphere. There the structure of the Klauder- Perelomov coherent
states associated to the $su(1, 1)$ algebra is considered.

On the other hand, recently, significant progress was achieved in
development of the nonlinear coherent states( NLCSs) or f-CSs, $|z,
{f}\rangle$, which are described by a specified nonlinearity
function ${f}(\hat{N})$ \cite{Matos2} and linked to nonlinear or
deformed algebras rather than the Lie algebras. NLCSs are defined as
the right-hand eigenstates of the non-Hermitian and deformed
annihilation operator ${f}(\hat{N})\hat{a}$, i.e.
\begin{eqnarray}
&&\hspace{-15mm}{f}(\hat{N})\hat{a}|z, {f}\rangle=z|z,
{f}\rangle\nonumber\end{eqnarray},or by applying the generalized
displacement operator on the ground state:
\begin{eqnarray}
&&\hspace{-15mm}|z,f\rangle=e^{\frac{z}{{f}(\hat{N}-1)}{\hat{a}}^{\dag}}|0\rangle\nonumber,\end{eqnarray}
where $\hat{a}$ and $\hat{a}^{\dag}$ are the annihilation and
creation operators of the harmonic oscillator, ${f}(\hat{N})$ is an
operator-valued function of the number operator $\hat{N} =
{\hat{a}}^{\dag}\hat{a}$. Obviously, $|z, f\rangle$ becomes the
canonical coherent state when ${f}(\hat{N})= 1$. In fact the nature
of the nonlinearity depends on the choice of the function
$f(\hat{N})$ \cite{Manko, Shanta}. These states may appear as
stationary states of the center-of mass motion of a trapped ion
\cite{Matos2}, and exhibit nonclassical features such as quadrature
squeezing, sub-Poissonian statistics, anti-bunching as well as
self-splitting effects \cite{Kis, Agarwal, Sivakumar, dehghani}.

Thus this work is devoted to analysis and pay more attention on the
possibility of construction of DNSs which can be raised
systematically, in an algebraic way, with the help of the
Barut-Girardello type of displacement operators,
$\mathcal{D}_{BG}(\mathfrak{z})$, instead of the Klauder- Perelomov
counterpart, acting on arbitrary Fock states $|m,\lambda\rangle$. As
will be shown, they illustrate more pronounced and controllable
statistical as well as nonclassical properties than the well known
DNSs and this is main reason to study them, carefully. In order to
justify the effectiveness of this technique, wide range of
non-classical quantum states associated to Lie group $SU(1, 1)$ and
corresponding with the Calogero-Sutherland model will be
constructed. Because of the two-particle Calogero-Sutherland model
has attracted considerable interests \cite{Calogero, Agarwal1,
dehghani1} and enjoys the $su(1,1)$ dynamic symmetry
\cite{Perelemov1, Perelemov2}, too. It is of great interest in
quantum optics because it can characterize many kinds of quantum
optical systems \cite{Barut, Perelomov, Zhang}. In particular, the
bosonic realization of $su(1,1)$ describes the degenerate and
non-degenerate parametric amplifiers \cite{WODKIEWICZ}. We will
prove that they are actually new kinds of {\em{photon-added type of
Barut-Girardello coherent states}} (PABGCSs),
$||{\mathfrak{z}},m\rangle_{\lambda}$, because of these states are
emerged through an iterated action of a creation operator,
${J}_{+}^{\lambda}$, on {\em{nonlinear Barut-Girardello coherent
states}} ( NBGCSs), $|{\mathfrak{z}},
\mathfrak{f}_{m}\rangle_{\lambda}$. While the latter are obtained
through an action of $m-$deformed Barut-Girardello displacement
operator, $\mathcal{D}^{m}_{BG}(\mathfrak{z})$, on vacuum state,
$|0,\lambda\rangle$. The reason for using the word "nonlinear" is
that, $|{\mathfrak{z}}, \mathfrak{f}_{m}\rangle_{\lambda}$ include
nonclassical features such as squeezing, anti-bunching effects and
sub-Poissonian statistics. The outcome of the above discussion can
be summarized in the following commutative diagram
\begin{eqnarray}&&\hspace{-12mm}\mathcal{D}^{m}_{BG}(\mathfrak{z})\nonumber\\
&&\hspace{-30mm}|0,\lambda\rangle\hspace{10mm}\Rightarrow\hspace{15mm}|{\mathfrak{z}}, \mathfrak{f}_{m}\rangle_{\lambda}\nonumber\\
&&\hspace{-42mm}\vspace{15mm}\nonumber\\
&&\hspace{-42mm}\left({J}_{+}^{\lambda}\right)^m\hspace{3mm}{{\Downarrow}}
\hspace{42mm}{{\Downarrow}}\left({J}_{+}^{\lambda}\right)^m\nonumber\\
&&\hspace{-12mm}\mathcal{D}_{BG}(\mathfrak{z})\nonumber\\
&&\hspace{-30mm}|m,\lambda\rangle\hspace{10mm}\Rightarrow\hspace{15mm}||{\mathfrak{z}},m\rangle_{\lambda}\nonumber\end{eqnarray}

The paper is organized as follows. By reviewing some aspects of the
two-particle Calogero-Sutherland model in section 2, we introduce
new kind of PABGCSs $||\mathfrak{z},m\rangle_{\lambda}$ and show
their over completeness and resolution to the identity properties,
will brought in section 4 deal with the general study of non
classicality of these states. Section 3 includes detailed studies on
NBGCSs and their properties. There to realize the resolution of the
identity condition, we have found the positive definite measures on
the complex plane and also their tendency to the well-known
Barut-Girardello coherent states( BGCSs) \cite{dehghani1} is
reviewed. Some interesting features are found. For instance, we have
shown that they can be considered as an eigenstate of certain
annihilation operators and can be interpreted as NLCSs with special
nonlinearity functions. Furthermore, it has been discussed in detail
that they evolve in time as like as the canonical coherent states.
In other words NBGCS possess the temporal stability property, too.
We conclude in section 5.\newpage
\section{Review and Construction}
In our work we consider dynamics of a single bosonic mode, described
by the Calegero-Sutherland Hamiltonian $H^{\lambda}$ on the
half-line $x$
\begin{eqnarray}
&&\hspace{-1.5cm}H^{\lambda}=\frac{1}{2}\left[-\frac{d^2}{dx^2}+x^2+\frac{\lambda(\lambda-1)}{x^2}\right].\end{eqnarray}
Here, the simple an-harmonic term $\frac{\lambda(\lambda-1)}{x^2}$
refers to the Goldman-Krivchenkov potential \cite{sasaki}. In Refs.
\cite{ dehghani1, Hall}, it has been shown that the second-order
differential operators
\begin{eqnarray}
&&\hspace{-1.5cm} {J}_{\pm}^{\lambda}:=\frac{1}{4}\left[\left(x\mp\frac{d}{dx}\right)^2-\frac{\lambda(\lambda-1)}{x^2}\right],\\
&&\hspace{-1.5cm}{J}_{3}^{\lambda}:=\frac{H^{\lambda}}{2},\end{eqnarray}
satisfy the standard commutation relations of $su(1,1)$ Lie algebra
as follows
\begin{eqnarray}
&&\hspace{-1.5cm}\left[{J}_{+}^{\lambda},{J}_{-}^{\lambda}\right]=-2{J}_{3}^{\lambda},
\hspace{20mm}\left[{J}_{3}^{\lambda},{J}_{\pm}^{\lambda}\right]=\pm{J}_{\pm}^{\lambda}.
\end{eqnarray}
Also, product the unitary and positive-integer IREP of $su(1,1)$ Lie
algebra as
\begin{eqnarray}&&\hspace{-15mm}{J}_{+}^{\lambda}|n-1,\lambda\rangle=\sqrt{n\left(n+\lambda-\frac{1}{2}\right)}|n,\lambda\rangle,\\
&&\hspace{-15mm}{J}_{-}^{\lambda}|n,\lambda\rangle=\sqrt{n\left(n+\lambda-\frac{1}{2}\right)}|n-1,\lambda\rangle,\\
&&\hspace{-15mm}{J}_{3}^{\lambda}|n,\lambda\rangle=\left(n+\frac{\lambda}{2}+\frac{1}{4}\right)|n,\lambda\rangle.\end{eqnarray}
We assume that the set of states described above form complete and
orthonormal basis
of an infinite dimensional Hilbert space, i.e. \begin{eqnarray}&&\hspace{-15mm}\mathcal{H}^{\lambda}:={\mathrm{span}}\{|n,\lambda\rangle|\langle n, \lambda|m, \lambda\rangle=\delta_{nm}\}|_{n=0}^{\infty},\nonumber\\
&&\hspace{-15mm}\langle
x|n,\lambda\rangle:=(-1)^n\sqrt{\frac{2\Gamma(n+1)}{\Gamma(n+\lambda+\frac{1}{2})}}x^{\lambda}
e^{-\frac{x^2}{2}}L_{n}^{\lambda-\frac{1}{2}}(x^2),
\hspace{4mm}\lambda >\frac{-1}{2},\end{eqnarray} where
$L_{n}^{\lambda-\frac{1}{2}}(x)$ denotes the associated Laguerre
functions \cite{Gradshteyn}. Along with the orthogonality of the
associated Laguerre polynomials, the orthogonality relation of the
basis of $\mathcal{H}^{\lambda}$ reads
\begin{eqnarray}&&\hspace{-15mm}\langle n, \lambda|m, \lambda\rangle:=\frac{2n!}{\Gamma(n+\lambda+\frac{1}{2})}\int_{0}^{\infty}x^{2\lambda}e^{-x^2}L_{n}^{\lambda-\frac{1}{2}}(x^2)
L_{m}^{\lambda-\frac{1}{2}}(x^2)dx=\delta_{nm}.\end{eqnarray} It is
useful to stress that the two operators ${J}_{+}^{\lambda}$ and
${J}_{-}^{\lambda}$ are Hermitian conjugate of each others with
respect to the inner product (9) and ${J}_{3}^{\lambda}$ is
self-adjoint operator, too. We commence by establishing our
formalism by collecting some well known facts about BGCSs, which are
defined as the action of the B-G displacement operator
$\mathcal{D}_{BG}(\mathfrak{z})$ on a vacuum state
\begin{eqnarray}
&&\hspace{-15mm}|\mathfrak{z}, 0\rangle_{\lambda}:\equiv\mathcal{D}_{BG}(\mathfrak{z})|0,\lambda\rangle\\
&&\hspace{-15mm}\mathcal{D}_{BG}(\mathfrak{z}) := e^{\frac{\mathfrak{z}}{\hat{N}+\lambda-\frac{1}{2}}{J}_{+}^{\lambda}}\\
&&\hspace{-15mm}\hat{N}: =
{J}_{3}^{\lambda}-\frac{\lambda}{2}-\frac{1}{4},\hspace{4mm}\hat{N}|n,\lambda\rangle=
n|n,\lambda\rangle\end{eqnarray} and, introduce the following
modifications for a deformed setting
\begin{eqnarray}&&\hspace{-15mm}||\mathfrak{z}, m\rangle_{\lambda}:={M_{m, \lambda}}^{-\frac{1}{2}}(|\mathfrak{z}|)\mathcal{D}_{BG}(\mathfrak{z})|m,\lambda\rangle,\end{eqnarray}
where $\mathfrak{z}$$(= |\mathfrak{z}|e^{i\varphi},\hspace{2mm}
0\leq|\mathfrak{z}|,\hspace{2mm} 0\leq\varphi\leq2\pi)$ is the
coherence parameter. Clearly, the normalized states $||\mathfrak{z},
m\rangle_{\lambda}$ become standard Barut-Girardello coherent states
for the Calegero-Sutherland model,
$|z\rangle^{\lambda}_{BG}\left(=|\mathfrak{z},
0\rangle_{\lambda}\right)$ (Eq. (6) in Ref. \cite{dehghani1}), while
$m$ tends to zero. Putting
\begin{eqnarray}&&\hspace{-15mm}|m,\lambda\rangle=
\frac{\left({J}_{+}^{\lambda}\right)^{m}}{\sqrt{m!(\lambda+\frac{1}{2})_{m}}}|0,\lambda\rangle\end{eqnarray}
into (13), we obtain
\begin{eqnarray}&&\hspace{-15mm}||{\bf{\mathfrak{z}}}, m\rangle_{\lambda}=\frac{e^{\frac{{\mathfrak{z}}}{\hat{N}+\lambda-\frac{1}{2}}{J}_{+}^{\lambda}}\left({J}_{+}^{\lambda}\right)^{m}}{\sqrt{m!(\lambda+\frac{1}{2})_{m}M_{m, \lambda}(|{\bf{\mathfrak{z}}}|)}}|0,\lambda\rangle,\end{eqnarray}
where $(a)_{m}=\frac{\Gamma(a+m)}{\Gamma(a)}$ denotes the
Pochammer's notation. The above expression after restoring the
formula
\begin{eqnarray}&&\hspace{-15mm}
\frac{\bf{\mathfrak{z}}}{\hat{N}+\lambda-\frac{1}{2}}{J}_{+}^{\lambda}={J}_{+}^{\lambda}\frac{\mathfrak{z}}{\hat{N}+\lambda-\frac{1}{2}+1},\nonumber\\
&&\hspace{-21mm}\frac{\bf{\mathfrak{z}}}{\hat{N}+\lambda-\frac{1}{2}}\left({J}_{+}^{\lambda}\right)^{2}=\left({J}_{+}^{\lambda}\right)^{2}\frac{\mathfrak{z}}{\hat{N}+\lambda-\frac{1}{2}+2},\nonumber\\
&&\hspace{10mm}\vdots\nonumber\\
&&\hspace{9mm}\Downarrow\nonumber\\
&&\hspace{-23mm}e^{\frac{\bf{\mathfrak{z}}}{\hat{N}+\lambda-\frac{1}{2}}{J}_{+}^{\lambda}}\left({J}_{+}^{\lambda}\right)^{m}=\left({J}_{+}^{\lambda}\right)^{m}e^{\frac{\bf{\mathfrak{z}}}{\hat{N}+\lambda-\frac{1}{2}+m}{J}_{+}^{\lambda}},\end{eqnarray}
yields
\begin{eqnarray}&&\hspace{-28mm}||{\bf{\mathfrak{z}}}, m\rangle_{\lambda}=\frac{\left({J}_{+}^{\lambda}\right)^{m}}{\sqrt{m!(\lambda+\frac{1}{2})_{m}{M_{m, \lambda}}(|\bf{\mathfrak{z}}|)}}e^{\frac{{\bf{\mathfrak{z}}}}{\hat{N}+\lambda-\frac{1}{2}+m}{J}_{+}^{\lambda}}|0,\lambda\rangle\nonumber\\
&&\hspace{-15mm}=\frac{\left({J}_{+}^{\lambda}\right)^{m}}{\sqrt{m!\left(\lambda+\frac{1}{2}\right)_{m}\frac{M_{m,
\lambda}(|{\bf{\mathfrak{z}}}|)}{\mathfrak{M}_{m,
\lambda}(|{\bf{\mathfrak{z}}}|)}}}|{\mathfrak{z}},
\mathfrak{f}_{m}\rangle_{\lambda},\end{eqnarray} here
$\mathfrak{f}_{m}$, that will be determined later, is an
operator-valued function of the number operator $\hat{N}$ and
$\mathfrak{M}_{m, \lambda}(|{\bf{\mathfrak{z}}}|)$ is chosen so that
$|{\mathfrak{z}}, \mathfrak{f}_{m}\rangle_{\lambda}$ is normalized,
i.e. $_{\lambda}\langle {\mathfrak{z}},
\mathfrak{f}_{m}|{\mathfrak{z}},
\mathfrak{f}_{m}\rangle_{\lambda}=1$. At this stage, we will
postpone investigation of the properties of $||{\bf{\mathfrak{z}}},
m\rangle_{\lambda}$ up to next section and consciously concentrate
to solve a problem that how the states $|{\mathfrak{z}},
\mathfrak{f}_{m}\rangle_{\lambda}$ can be recognized as nonlinear
type of coherent states.

\section{{\em{NBGCSs}} And It's Properties}
In this section we want to inform that the states $|{\mathfrak{z}}, \mathfrak{f}_{m}\rangle_{\lambda}$ were introduced above as {\em{NBGCSs}}, can be categorized as special class of nonlinear coherent states. For this reason we will set up detailed studies on statistical properties of them. Proportional nonlinear function associated to {\em{NBGCSs}} are introduced, also to analyze their statistical behavior some of the characters including the second-order correlation function, Mandel's parameter and squeezing factor are computed.\\\\
$\diamondsuit$ {\bf{\em{NBGCS}}}\\
Let $|{\mathfrak{z}}, \mathfrak{f}_{m}\rangle_{\lambda}$ denotes the
states generated by {\em{m-}}deformed Barut-Girardello displacement
operator $\mathcal{D}^{m}_{BG}(\mathfrak{z})$, i.e.
\begin{eqnarray}&&\hspace{-18mm}|{\mathfrak{z}}, \mathfrak{f}_{m}\rangle_{\lambda}:=\mathfrak{M}_{m, \lambda}^{-\frac{1}{2}}(|{\bf{\mathfrak{z}}}|)\mathcal{D}^{m}_{BG}(\mathfrak{z})|0,\lambda\rangle,\hspace{5mm} m\in N_{0},\\
&&\hspace{-18mm}\mathcal{D}^{m}_{BG}(\mathfrak{z}):=
e^{\frac{{\bf{\mathfrak{z}}}}{\hat{N}+\lambda-\frac{1}{2}+m}{J}_{+}^{\lambda}},\hspace{22mm}\lim_{m\rightarrow
0}\mathcal{D}^{m}_{BG}(\mathfrak{z})=\mathcal{D}_{BG}(\mathfrak{z}).\end{eqnarray}
Inserting (14) into (18) and using the identity
\begin{eqnarray}&&\hspace{-18mm}\left({\frac{1}{\hat{N}+\lambda-\frac{1}{2}+m}{J}_{+}^{\lambda}}\right)^n=
\left({J}_{+}^{\lambda}\right)^{n}\frac{1}{\left(\hat{N}+\lambda+\frac{1}{2}+m\right)_{n}},\end{eqnarray}
we obtain
\begin{eqnarray}&&\hspace{-18mm}|{\mathfrak{z}}, \mathfrak{f}_{m}\rangle_{\lambda}=\mathfrak{M}_{m, \lambda}^{-\frac{1}{2}}(|{\bf{\mathfrak{z}}}|)
\sum_{n=0}^{\infty}{\frac{\mathfrak{z}^{n}}{\left(\lambda+\frac{1}{2}+m\right)_{n}}\sqrt{\frac{\left(\lambda+\frac{1}{2}\right)_{n}}{n!}}}\hspace{2mm}|n,\lambda\rangle.\end{eqnarray}
Due to the orthogonality relation (9) it follows that overlapping of
two different kinds of these normalized states must be
nonorthogonal, if $m'\neq m$ and $\mathfrak{z'}\neq \mathfrak{z}$,
i.e.
\begin{eqnarray}&&\hspace{-15mm}{_\lambda}\langle {\mathfrak{z'}}, \mathfrak{f}_{m'}|{\mathfrak{z}}, \mathfrak{f}_{m}\rangle_{\lambda}=\frac{_{1}F_{2}\left(\left[\lambda+\frac{1}{2}\right],\left[\lambda+\frac{1}{2}+m' , \lambda+\frac{1}{2}+m\right],\overline{\mathfrak{z'}}\mathfrak{z}\right)}{\sqrt{{\mathfrak{M}_{m'}^{\lambda}}(|z'|){\mathfrak{M}_{m}^{\lambda}}(|z|)}}.\end{eqnarray}
Then, ${\mathfrak{M}_{m, \lambda}}(|\mathfrak{z}|)$ can be
calculated to be taken as
\begin{eqnarray}&&\hspace{-15mm}{\mathfrak{M}_{m, \lambda}}(|\mathfrak{z}|)=_{1}F_{2}\left(\left[\lambda+\frac{1}{2}\right],\left[\lambda+\frac{1}{2}+m , \lambda+\frac{1}{2}+m\right],|\mathfrak{z}|^2\right).\end{eqnarray}
From the completeness relation of the Fock space states,
straightforwardly, resolution of the identity
\begin{eqnarray}&&\hspace{-5mm}\oint_{\mathbb{C}(\mathfrak{z})}{|{\mathfrak{z}}, \mathfrak{f}_{m}\rangle_{\lambda}\hspace{1mm}{_{\lambda}\langle {\mathfrak{z}}, \mathfrak{f}_{m}|}}d\mu_{m, \lambda}(|\mathfrak{z}|)=I^{\lambda}=\sum_{n=0}^{\infty}{|n,\lambda\rangle\langle n, \lambda|},\end{eqnarray}
is realized for the states $|{\mathfrak{z}},
\mathfrak{f}_{m}\rangle_{\lambda}$ with respect to the appropriate
measure, $d\mu_{m, \lambda}(|\mathfrak{z}|) := \mathfrak{K}_{m,
\lambda}(|\mathfrak{z}|) \frac{d|\mathfrak{z}|^{2}}{2}d\varphi$,
that relates it to the Meijer{'}s G-function( see $\frac{7-811}{4}$
in \cite{Gradshteyn}):
\begin{eqnarray}&&\hspace{-25mm}\mathfrak{K}_{m, \lambda}(|\mathfrak{z}|)
=\frac{{\mathfrak{M}_{m, \lambda}}(|\mathfrak{z}|)}{\pi
\left(\lambda+\frac{1}{2}\right)_{m}\Gamma(\lambda+\frac{1}{2}+m)}\mathbf{G}^{3
1}_{2 4}\left(|\mathfrak{z}|^2 {\mid} \hspace{1mm}^{0,\hspace{1mm}
\lambda-\frac{1}{2}}_{0,\hspace{1mm}
\lambda-\frac{1}{2}+m,\hspace{1mm}
\lambda-\frac{1}{2}+m,\hspace{1mm} 0}\right),
\hspace{1mm}\lambda-\frac{1}{2}\in 2\mathbb{N}_{0}\end{eqnarray} It
is worth to mention that $\mathfrak{K}_{m=0,
\lambda}(|\mathfrak{z}|)$ reduces to the well known positive
definite measure of the standard Barut-Girardello coherent states,
i.e.
\begin{eqnarray}&&\hspace{1mm}\lim_{m\rightarrow 0}\mathfrak{K}_{m, \lambda}(|\mathfrak{z}|)=\frac{2}{\pi}I_{\lambda-\frac{1}{2}}(2|{\mathfrak{z}}|)K_{\lambda-\frac{1}{2}}(2|{\mathfrak{z}}|),\nonumber\end{eqnarray}
where $I_{a}(x)$ and $K_{b}(x)$ refer to the modified Bessel functions of the first and second kinds, respectively( see Figure 1).\\\\
\begin{figure}
\begin{center}
\epsfig{figure=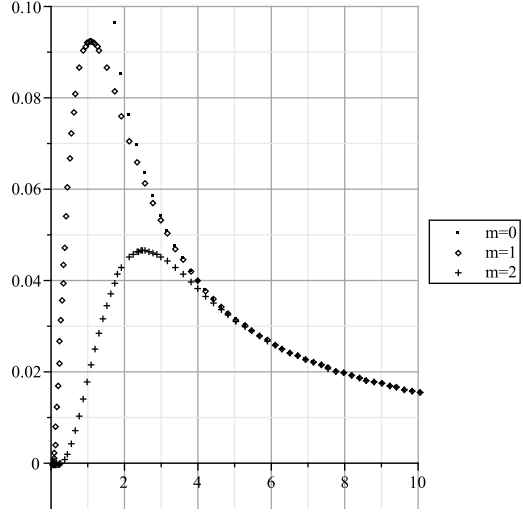,width=7cm}
\end{center}
\caption{\footnotesize Plots of the non-oscillating measures
$\mathfrak{K}_{m, \lambda}(|\mathfrak{z}|)$ in terms of
$|\mathfrak{z}|$ for different values of $m$ and
$\lambda=\frac{1}{2}$. The dotted curve corresponds to the standard
B-G coherent states.}
\end{figure}
$\diamondsuit$ {{\bf\em{Time Evolution }}}\\
Based on the relations (3) and (8), we have
\begin{eqnarray}
&&\hspace{-14mm}H^{\lambda}|n,\lambda\rangle=(2n+\lambda+\frac{1}{2})|n,\lambda\rangle.\end{eqnarray}
Then the states (21) evolve in time as
\begin{eqnarray}
&&\hspace{-14mm}e^{-itH^{\lambda}}|{\mathfrak{z}},
\mathfrak{f}_{m}\rangle_{\lambda}=e^{-it\left(\lambda+\frac{1}{2}\right)}\mathfrak{M}_{m,
\lambda}^{-\frac{1}{2}}(|{\bf{\mathfrak{z}}}|)\sum_{n=0}^{\infty}{\frac{\left(\mathfrak{z}e^{-i2t}\right)^{n}}{\left(\lambda+
\frac{1}{2}+m\right)_{n}}\sqrt{\frac{\left(\lambda+\frac{1}{2}\right)_{n}}{n!}}}\hspace{2mm}|n,\lambda\rangle\nonumber\\
&&\hspace{6mm}=e^{-it\left(\lambda+\frac{1}{2}\right)}\left|\mathfrak{z}e^{-i2t},\mathfrak{f}_{m}\right\rangle_{\lambda}.\end{eqnarray}
In fact, these states are temporally stable.\\\\
$\diamondsuit$ {{\bf\em{Coordinate Representation of}} $|{\mathfrak{z}}, \mathfrak{f}_{n}\rangle_{\lambda}$}\\
Based on a new expression of the Laguerre polynomials as an
operator-valued function given in \cite{Cessa}
\begin{eqnarray}&&\hspace{-15mm}L_{n}^{\alpha}(y) =
\frac{1}{n!}y^{-\alpha}\left(\frac{d}{dy}-1\right)^n
y^{n+\alpha},\end{eqnarray} also, according to Eqs. (8) and (21) we
have
\begin{eqnarray}&&\hspace{-7mm}\langle x|{\mathfrak{z}}, {\mathfrak{f}_{m}}\rangle_{\lambda}=\sqrt{\frac{2}{\Gamma(\lambda+\frac{1}{2}){\mathfrak{M}}_{m, \lambda}}}(|{\bf{\mathfrak{z}}}|)\sum_{n=0}^{\infty}{\frac{(-{\mathfrak{z}})^{n}}{\left(\lambda+\frac{1}{2}+m\right)_{n}}}\hspace{2mm}x^{\lambda}e^{-\frac{x^2}{2}}
L_{n}^{\lambda-\frac{1}{2}}(x^2)\nonumber\\
&&\hspace{-7mm}=\sqrt{\frac{2}{\Gamma(\lambda+\frac{1}{2}){\mathfrak{M}}_{m,
\lambda}}}(|{\bf{\mathfrak{z}}}|)\sum_{n=0}^{\infty}{\frac{(-{\mathfrak{z}})^{n}}{n!\left(\lambda+\frac{1}{2}+m\right)_{n}}}
e^{-\frac{y}{2}}y^{\frac{1-\lambda}{2}}\left(\frac{d}{dy}-1\right)^n
y^{n+\lambda-\frac{1}{2}}|_{y=x^2}\nonumber.\end{eqnarray} If we
resort
\begin{eqnarray}&&\hspace{-15mm}
\left(\frac{d}{dy}-1\right)^n
y^{n}=\left(y\frac{d}{dy}+n-y\right)...\left(y\frac{d}{dy}+1-y\right)=\left(y\frac{d}{dy}-y+1\right)_{n},\nonumber\end{eqnarray}
it becomes
\begin{eqnarray}&&\hspace{-7mm}\langle x|{\mathfrak{z}}, {\mathfrak{f}_{m}}\rangle_{\lambda}=\sqrt{{\frac{2e^{-{y}}y^{{1-\lambda}}}{\Gamma(\lambda+\frac{1}{2}){\mathfrak{M}_{m, \lambda}}(|{\bf{\mathfrak{z}}}|)}}}\sum_{n=0}^{\infty}{\frac{\left(y\frac{d}{dy}-y+1\right)_{n}(-{\mathfrak{z}})^{n}}{n!\left(\lambda+\frac{1}{2}+m\right)_{n}}}
y^{\lambda-\frac{1}{2}}|_{y=x^2}\nonumber\\
&&\hspace{-9mm}=\sqrt{{\frac{2e^{-{y}}y^{{1-\lambda}}}{\Gamma(\lambda+\frac{1}{2})}}}\hspace{1mm}
\frac{{_{1}F_{1}\left(\left[y\frac{d}{dy}-y+1\right],
\left[\lambda+\frac{1}{2}+m\right],-\mathfrak{z}\right)}y^{\lambda-\frac{1}{2}}}{\sqrt{_{1}F_{2}\left(\left[\lambda+\frac{1}{2}\right],\left[\lambda+\frac{1}{2}+m
,
\lambda+\frac{1}{2}+m\right],|\mathfrak{z}|^2\right)}}|_{y=x^2},\end{eqnarray}
where $_{1}F_{1}$ is the (Kummer's) confluent hypergeometric
function which corresponds to the special case $u= v= 1$ of the
generalized hypergeometric function $_{u}F_{v}$ (with $u$ numerator
and $v$ denominator parameters). For instance, the explicit compact
forms of $|{\mathfrak{z}}, \mathfrak{f}_{0}\rangle_{\lambda}$
reduces to the standard Barut-Girardello coherent states
\cite{dehghani1}
\begin{eqnarray}
&&\hspace{-14mm}\langle x|{\mathfrak{z}},
\mathfrak{f}_{0}\rangle_{\lambda}=\sqrt{2x}\left(-\frac{\mathfrak{z}}{|\mathfrak{z}|}\right)^{\frac{1}{4}-\frac{\lambda}{2}}
\frac{J_{\lambda-\frac{1}{2}}(2ix\sqrt{\mathfrak{z}})}{\sqrt{I_{\lambda-\frac{1}{2}}(2|\mathfrak{z}|)}}e^{-\mathfrak{z}-\frac{x^2}{2}}\end{eqnarray}
$\diamondsuit$ {{\bf{\em{Nonlinearity function}}}}\\
Using the relations (6) and (21), one can show that
\begin{eqnarray}&&\hspace{-10mm}{J}_{-}^{\lambda}|{{\mathfrak{z}}}, {\mathfrak{f}_{m}}\rangle_{\lambda}={\mathfrak{M}_{m,
\lambda}}^{-\frac{1}{2}}(|{\bf{\mathfrak{z}}}|)\sum_{n=0}^{\infty}{\left[{\mathfrak{z}}\frac{\lambda+\frac{1}{2}+n}{\lambda+\frac{1}{2}+m+n}\right]
\frac{{\mathfrak{z}}^{n}}{\left(\lambda+\frac{1}{2}+m\right)_{n}}\sqrt{\frac{\left(\lambda+\frac{1}{2}\right)_{n}}{n!}}}\hspace{2mm}|n,\lambda\rangle\nonumber\\
&&\hspace{55mm}\Downarrow\nonumber\\
&&\hspace{25mm}\left(1+\frac{m}{\hat{N}+\lambda+\frac{1}{2}}\right){J}_{-}^{\lambda}|{{\mathfrak{z}}},
{\mathfrak{f}_{m}}\rangle_{\lambda}={\mathfrak{z}}|{{\mathfrak{z}}},
{\mathfrak{f}_{m}}\rangle_{\lambda}.\end{eqnarray} This formula
specifies NBGCSs as new class of nonlinear coherent states with
characterized nonlinearity functions, $
\mathfrak{f}_{m}\left(=1+\frac{m}{\hat{N}+\lambda+\frac{1}{2}}\right)$,
which tend to unity for $m= 0$.\newpage
$\diamondsuit${{\bf\em{Anti-bunching effect and sub-Poissonian statistics}}}\\
Now we are in a position to study the anti-bunching effect as well
as the statistics of {\em{NBGCS}}s. We introduce the second-order
correlation function for these states
\begin{eqnarray}
&&\hspace{-14mm}\left(g^{(2)}\right)_{m}^{\lambda}(|\mathfrak{z}|)=
\frac{\langle {\hat{N}}^2\rangle_{m}^{\lambda}-{\langle{\hat{N}
\rangle}_{m}^{\lambda}}}{{\langle{\hat{N}
\rangle}_{m}^{\lambda}}^{2}}.\end{eqnarray} Furthermore, the
inherent statistical properties of the GNCSs follows also from
calculating the Mandel parameter \footnote{A state for which
$\mathfrak{Q}_{m}^{\lambda}(|\mathfrak{z}|)> 0$ (or
$\left(g^{(2)}\right)_{m}^{\lambda}(|\mathfrak{z}|)> 1$) is called
super-Poissonian (bunching effect), if $\mathfrak{Q} = 0$ (or
$g^{(2)}= 1$) the state is called Poissonian, while a state for
which $\mathfrak{Q}< 0$ (or $g^{(2)}< 1$) is, also, called
sub-Poissonian (antibunching effect).}
\begin{eqnarray}
&&\hspace{-14mm}\mathfrak{Q}_{m}^{\lambda}(|\mathfrak{z}|)=
{\langle{\hat{N}
\rangle}_{m}^{\lambda}}\left[\left(g^{(2)}\right)_{m}^{\lambda}(|\mathfrak{z}|)-1\right].
\end{eqnarray}
In order to find the function
$\left(g^{(2)}\right)_{m}^{\lambda}(|\mathfrak{z}|)$, also Mandel's
parameter $\mathfrak{Q}_{m}^{\lambda}(|\mathfrak{z}|)$, let us begin
with the expectation values of the number operator $\hat{N}$ and
it's square in the basis of the Fock states $\left|n,
\lambda\right\rangle$
\begin{figure}
\begin{center}
\epsfig{figure=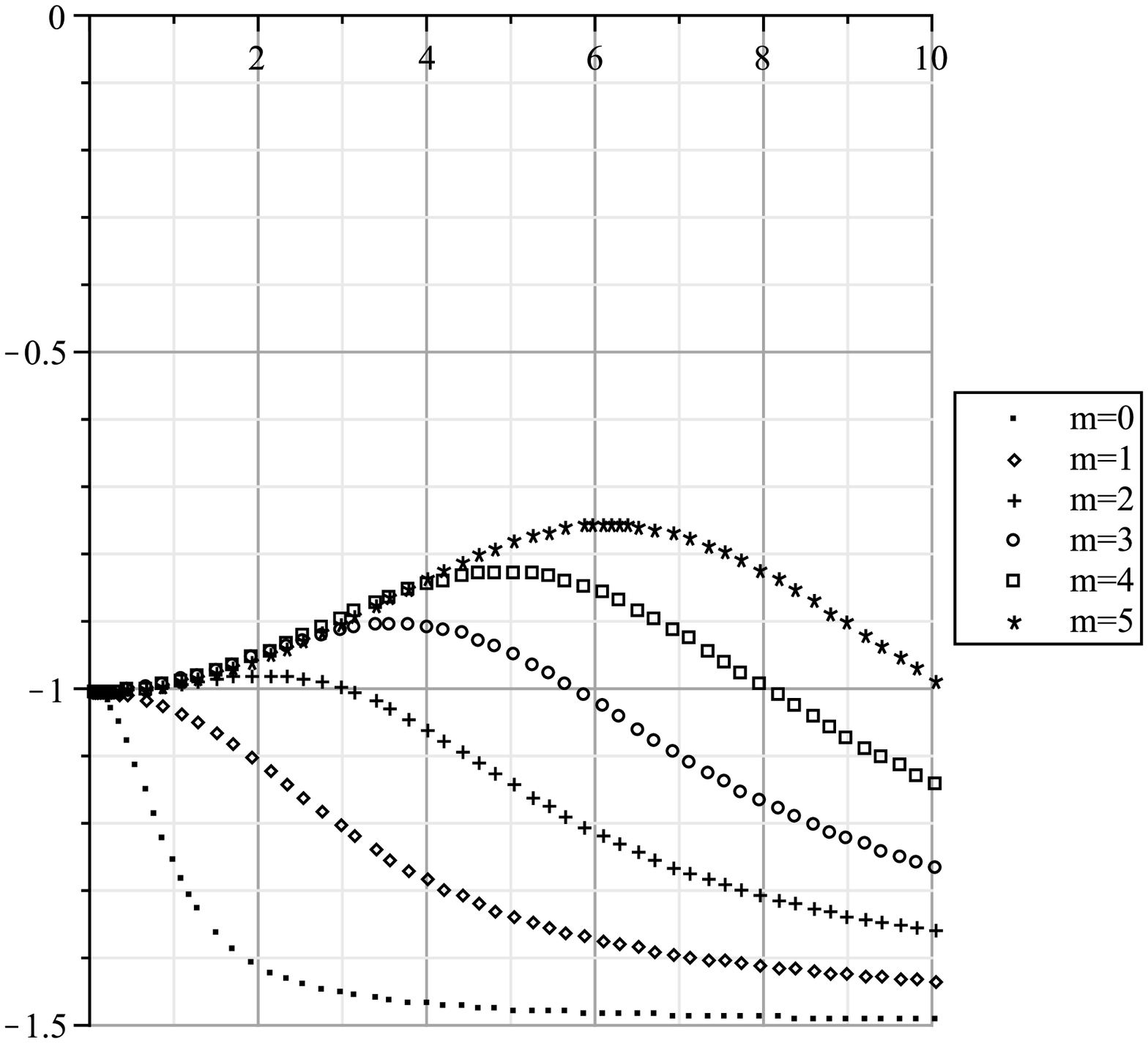,width=7cm}
\end{center}
\caption{\footnotesize Graphs of the Mandel's parameters
$\mathfrak{Q}_{m}^{\lambda}(|\mathfrak{z}|)$ in the NBGCSs, versus
$|\mathfrak{z}|$ for different values of $m$ and
$\lambda=\frac{1}{2}$. The dotted curve corresponds to the standard
B-G coherent states.}
\end{figure}
\begin{eqnarray}
&&\hspace{-7mm}{\langle{\hat{N}\rangle}_{m}^{\lambda}}={|\mathfrak{z}|^{2}}\left[\frac{\lambda+\frac{1}{2}}{(\lambda+\frac{1}{2}+m)^2}\right]
\frac{{_{1}F_{2}\left(\left[\lambda+\frac{3}{2}\right],\left[\lambda+\frac{3}{2}+m, \lambda+\frac{3}{2}+m\right],|\mathfrak{z}|^{2}\right)}}{_{1}F_{2}\left(\left[\lambda+\frac{1}{2}\right],\left[\lambda+\frac{1}{2}+m , \lambda+\frac{1}{2}+m\right],|\mathfrak{z}|^2\right)},\nonumber\\
&&\hspace{-7mm}\langle{\hat{N}}^2\rangle_{m}^{\lambda}={|\mathfrak{z}|^{4}}\left[\frac{\left(\lambda+\frac{1}{2}\right)
\left(\lambda+\frac{3}{2}\right)}{(\lambda+\frac{1}{2}+m)^2(\lambda+\frac{3}{2}+m)^2}\right]\nonumber\\
&&\hspace{40mm}\times\frac{{_{1}F_{2}\left(\left[\lambda+\frac{5}{2}\right],\left[\lambda+\frac{5}{2}+m,
\lambda+\frac{5}{2}+m\right],|\mathfrak{z}|^{2}\right)}}{_{1}F_{2}\left(\left[\lambda+\frac{1}{2}\right],\left[\lambda+\frac{1}{2}+m
,
\lambda+\frac{1}{2}+m\right],|\mathfrak{z}|^2\right)}.\nonumber\end{eqnarray}
Our calculations show that for any case $m\in {\mathbb{N}}_{0}$ as
well as $\lambda\in 2\mathbb{N}_{0}+\frac{1}{2}$, the Mandel's
parameters are really smaller than zero. In other words the
{\em{NBGCS}}s exhibit fully anti-bunching effects, or sub-Poissonian
statistics.
As shown in figure 2, $\mathfrak{Q}_{m}^{\lambda}(|\mathfrak{z}|)$ have been plotted in terms of $|\mathfrak{z}|$ for several values of $m$(= 0, 1, 2, 3, 4 and 5) where we choose $\lambda=\frac{1}{2}$.\\\\
$\diamondsuit$ {\bf{\em{$SU(1,1)$ squeezing}}}\\
We introduce two generalized Hermitian quadrature operators
$X_{1(2)}^{\lambda}$
\begin{eqnarray}
&&\hspace{-14mm}X_{1}^{\lambda}=\frac{{J}_{+}^{\lambda}+{J}_{-}^{\lambda}}{2},\hspace{4mm}X_{2}^{\lambda}=\frac{{J}_{-}^{\lambda}-{J}_{+}^{\lambda}}{2i},\end{eqnarray}
with the commutation relation $[X_{1}^{\lambda}, X_{2}^{\lambda}] =
i{J}_{3}^{\lambda}$. From this, the uncertainty condition for the
variances of the quadratures $X_{i}$ follow
\begin{eqnarray}
&&\hspace{-14mm}\langle(\Delta X_{1}^{\lambda})^2\rangle
\langle(\Delta X_{2}^{\lambda})^2\rangle \geq
\frac{|\langle{J}_{3}^{\lambda}\rangle|^{2}}{4},\end{eqnarray} where
$\langle(\Delta X_{i}^{\lambda})^2\rangle\left( =\langle{
X_{i}^{\lambda}}^2\rangle-\langle X_{i}^{\lambda}\rangle^2\right)$
have been expressed as
\begin{eqnarray}
&&\hspace{-14mm}\langle(\Delta X_{1(2)}^{\lambda})^2\rangle
=\frac{2\left\langle{J}_{+}^{\lambda}{J}_{-}^{\lambda}\right\rangle+2\left\langle{J}_{3}^{\lambda}\right\rangle\pm\left\langle{{J}_{+}^{\lambda}}^{2}+{{J}_{-}^{\lambda}}^{2}\right\rangle-
{\left\langle{{J}_{-}^{\lambda}}\pm{{J}_{+}^{\lambda}}\right\rangle}^{2}}{4}.\end{eqnarray}
Here, the angular brackets denote averaging over an arbitrary
normalizable state for which the mean values are well defined, i.e.
\begin{eqnarray}
&&\hspace{-14mm}\langle X_{i}\rangle={_{\lambda}\langle
\mathfrak{z}, \mathfrak{f}_{m}|}X_{i}|\mathfrak{z},
\mathfrak{f}_{m}\rangle_{\lambda}.\nonumber\end{eqnarray} For
instance, we have the relations
\begin{eqnarray}&&\hspace{-8mm}\left\langle{J}_{+}^{\lambda}\right\rangle=\overline{\left\langle{J}_{-}^{\lambda}\right\rangle}=
\overline{\mathfrak{z}}\left[\frac{\lambda+\frac{1}{2}}{\lambda+\frac{1}{2}+m}\right]
\frac{{_{1}F_{2}\left(\left[\lambda+\frac{3}{2}\right],\left[\lambda+\frac{3}{2}+m,
\lambda+\frac{1}{2}+m\right],|\mathfrak{z}|^{2}\right)}}{_{1}F_{2}\left(\left[\lambda+\frac{1}{2}\right],\left[\lambda+\frac{1}{2}+m
,\lambda+\frac{1}{2}+m\right],|\mathfrak{z}|^2\right)},\nonumber\end{eqnarray}
\begin{eqnarray}&&\hspace{-8mm}\left\langle{{J}_{+}^{\lambda}}^{2}\right\rangle=\overline{\left\langle{{J}_{-}^{\lambda}}^{2}\right\rangle}=
{\overline{\mathfrak{z}}}^{2}\left[\frac{(\lambda+\frac{1}{2})(\lambda+\frac{3}{2})}{(\lambda+\frac{1}{2}+m)(\lambda+\frac{3}{2}+m)}\right]\nonumber\\
&&\hspace{42mm}\times\frac{{_{1}F_{2}\left(\left[\lambda+\frac{5}{2}\right],\left[\lambda+\frac{5}{2}+m, \lambda+\frac{1}{2}+m\right],|\mathfrak{z}|^{2}\right)}}{_{1}F_{2}\left(\left[\lambda+\frac{1}{2}\right],\left[\lambda+\frac{1}{2}+m , \lambda+\frac{1}{2}+m\right],|\mathfrak{z}|^2\right)},\nonumber\\
&&\hspace{-4mm}\left\langle{{J}_{+}^{\lambda}}{J}_{-}^{\lambda}\right\rangle=
{|\mathfrak{z}|^{2}}\left[\frac{(\lambda+\frac{1}{2})\Gamma(\lambda+\frac{1}{2}+m)}{\Gamma(\lambda+\frac{3}{2}+m)}\right]^2\nonumber\\
&&\hspace{4mm}\times
\frac{{_{2}F_{3}\left(\left[\lambda+\frac{3}{2}, \lambda+\frac{3}{2}\right],\left[\lambda+\frac{1}{2}, \lambda+\frac{3}{2}+m, \lambda+\frac{3}{2}+m\right],|\mathfrak{z}|^{2}\right)}}{_{1}F_{2}\left(\left[\lambda+\frac{1}{2}\right],\left[\lambda+\frac{1}{2}+m , \lambda+\frac{1}{2}+m\right],|\mathfrak{z}|^2\right)},\nonumber\\
&&\hspace{-4mm}\left\langle{J}_{3}^{\lambda}\right\rangle={|\mathfrak{z}|^{2}}\left[\frac{\lambda+\frac{1}{2}}{(\lambda+\frac{1}{2}+m)^2}\right]
\frac{{_{1}F_{2}\left(\left[\lambda+\frac{3}{2}\right],\left[\lambda+\frac{3}{2}+m, \lambda+\frac{3}{2}+m\right],|\mathfrak{z}|^{2}\right)}}{_{1}F_{2}\left(\left[\lambda+\frac{1}{2}\right],\left[\lambda+\frac{1}{2}+m , \lambda+\frac{1}{2}+m\right],|\mathfrak{z}|^2\right)}\nonumber\\
&&\hspace{24mm}+\frac{\lambda}{2}+\frac{1}{4}.\nonumber\end{eqnarray}
They result that, $\langle(\Delta X_{1(2)}^{\lambda})^2\rangle$ for
any value of $\lambda$, are efficiently dependent on the complex
variable $\mathfrak{z}$ and the deformation parameter $m$.
\begin{figure}
\begin{center}
\epsfig{figure=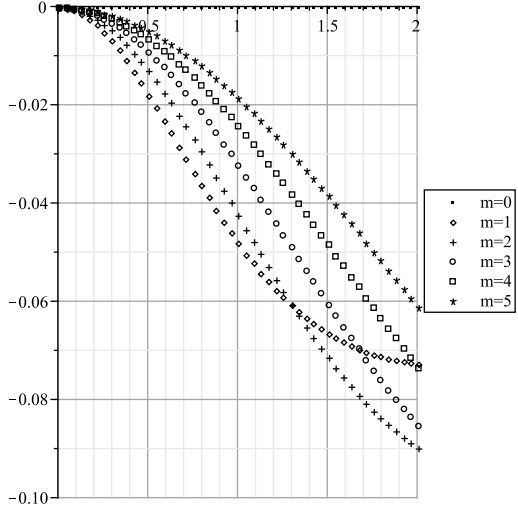,width=7cm}
\end{center}
\caption{\footnotesize Squeezing factors in the field quadratures
$X_{1}^{\lambda}$ in the NBGCSs, in terms of $|\mathfrak{z}|$ for
different values of $m$ and $\lambda=\frac{1}{2}$ while we choose
the phase $\varphi=\frac{\pi}{3}$. The dotted curve corresponds the
standard B-G coherent states( ${\mathcal{S}}^{m=0,
\lambda}_{1}=0$).}
\end{figure}

Following Walls (1983) we will say that the state is $SU(1,1)$
squeezed if the following condition is fulfilled \cite{Walls}
\begin{eqnarray}
&&\hspace{-14mm}\langle(\Delta X_{i}^{\lambda})^2\rangle <
\frac{|\langle{J}_{3}^{\lambda} \rangle|}{2},\hspace{4mm}
for\hspace{2mm} i=1 \hspace{2mm}or \hspace{2mm}2,\end{eqnarray} or,
with respect to the squeezing factor $\mathcal{S}^{m,
\lambda}_{i}(\mathfrak{z})$ as \cite{Buzek}
\begin{eqnarray}&&\hspace{-4mm}\mathcal{S}^{m, \lambda}_{i}(\mathfrak{z})\left(=2\frac{\langle(\Delta X_{i}^{\lambda})^2\rangle}{{|\langle{J}_{3}^{\lambda} \rangle|}}-1\right)< 0,\end{eqnarray}
however maximally squeezing is obtained for $\mathcal{S}^{m,
\lambda}_{i}(\mathfrak{z})=-1$. We illustrate in figure 3 squeezing
factors $\mathcal{S}^{m, \lambda}_{1}(\mathfrak{z})$ as functions of
$|\mathfrak{z}|$ for different values of $m$(= 0, 1, 2, 3, 4 and 5)
where we choose the phase $\varphi=\frac{\pi}{3}$ as well as
$\lambda=\frac{1}{2}$. They become really smaller than zero for any
values of $|\mathfrak{z}|$ and we find that by increasing $m$ the
degree of squeezing is enhanced. However we will loose the squeezing
in $X_{1}$ when $m$ reaches zero. In other words, for the case $m=0$
( standard BG coherent states) we would not expect to take squeezing
neither in $X_{1}$ nor in $X_{2}$ quadratures.\newpage
\section{{\em{PABGCS}}s}
According to (17), we introduce the state $||\bf{\mathfrak{z}},
m\rangle_{\lambda}$ defined by
\begin{eqnarray}&&\hspace{-28mm}||{\bf{\mathfrak{z}}}, m\rangle_{\lambda}=\frac{\left({J}_{+}^{\lambda}\right)^{m}}{\sqrt{m!\left(\lambda+\frac{1}{2}\right)_{m}\frac{M_{m, \lambda}(|{\bf{\mathfrak{z}}}|)}{\mathfrak{M}_{m, \lambda}(|{\bf{\mathfrak{z}}}|)}}}|{\mathfrak{z}}, \mathfrak{f}_{m}\rangle_{\lambda},\end{eqnarray}
where $|{\mathfrak{z}}, \mathfrak{f}_{m}\rangle_{\lambda}$ refer to
the NBGCSs. In the limit ${\mathfrak{z}}\rightarrow 0$ and $
m\rightarrow 0$ the states $||{\bf{\mathfrak{z}}},
m\rangle_{\lambda}$ reduce the Fock state and the standard
Barut-Girardello coherent state, respectively. Thus, it is a state
intermediate Fock state and BGCS, so we may call such states as {\bf
Photon Added Barut-Girardello Coherent States}.

Substituting of (21) into the right hand side of (17), leads to the
state $||\bf{\mathfrak{z}}, m\rangle_{\lambda}$ in terms of the Fock
states as
\begin{eqnarray}&&\hspace{-13mm}||\bf{\mathfrak{z}}, m\rangle_{\lambda}=\frac{\left({J}_{+}^{\lambda}\right)^{m}}{\sqrt{m!\left(\lambda+\frac{1}{2}\right)_{m}{M_{m, \lambda}(|\bf{\mathfrak{z}}|)}}}
\sum_{n=0}^{\infty}{\frac{\mathfrak{z}^{n}}{\left(\lambda+\frac{1}{2}+m\right)_{n}}
\sqrt{\frac{\left(\lambda+\frac{1}{2}\right)_{n}}{n!}}}\hspace{2mm}|n,\lambda\rangle\nonumber\\
&&\hspace{2mm}=\frac{\sum_{n=0}^{\infty}{\frac{\mathfrak{z}^{n}}{\left(\lambda+\frac{1}{2}+m\right)_{n}}
\sqrt{\frac{\left(\lambda+\frac{1}{2}\right)_{n}(n+m)!(n+\lambda-\frac{1}{2}+m)!}{n!^{2}(n+\lambda-\frac{1}{2})!}}}\hspace{2mm}|n+m,\lambda\rangle}{\sqrt{m!\left(\lambda+\frac{1}{2}\right)_{m}{M_{m, \lambda}(|\bf{\mathfrak{z}}|)}}}\nonumber\\
&&\hspace{2mm}=\frac{1}{\sqrt{{M_{m,
\lambda}(|\bf{\mathfrak{z}}|)}}}\sum_{n=0}^{\infty}{\frac{\mathfrak{z}^{n}}{n!}
\sqrt{\frac{(m+1)_{n}}{\left(\lambda+\frac{1}{2}+m\right)_{n}}}}\hspace{2mm}|n+m,\lambda\rangle.\end{eqnarray}
At the same time
\begin{eqnarray}&&\hspace{-28mm}{M_{m, \lambda}}(|\mathfrak{z}|)=_{1}F_{2}\left([m+1],\left[1, \lambda+\frac{1}{2}+m\right],|\mathfrak{z}|^2\right),\end{eqnarray}
where the latter results from the requirement
$_{\lambda}\langle\bf{\mathfrak{z}}, m||\bf{\mathfrak{z}},
m\rangle_{\lambda}=1$ and lead to the following non-vanishing scalar
products
\begin{eqnarray}&&\hspace{-12mm}_{\lambda}\langle\bf{\mathfrak{z}}, m'||\bf{\mathfrak{z}}, m\rangle_{\lambda}=\nonumber\\
&&\hspace{-15mm}\frac{\left[\frac{(1)_{m}\left(\lambda+\frac{1}{2}\right)_{m'}}{(1)_{m'}\left(\lambda+\frac{1}{2}\right)_{m}}\right]^{\frac{1}{2}}
{\mathfrak{z}}^{m-m'}\sum_{n=0}^{\infty}{\frac{|\mathfrak{z}|^{2n}}{n!}
{\frac{(m+1)_{n}}{\left(\lambda+\frac{1}{2}+m\right)_{n}(n+m-m')!}}}}{\sqrt{_{1}F_{2}\left([m'+1],\left[1, \lambda+\frac{1}{2}+m'\right],|\mathfrak{z}|^2\right)\hspace{1mm}_{1}F_{2}\left([m+1],\left[1, \lambda+\frac{1}{2}+m\right],|\mathfrak{z}|^2\right)}},\nonumber\\
&&\hspace{-12mm}_{\lambda}\langle\bf{\mathfrak{z'}}, m||\bf{\mathfrak{z}}, m\rangle_{\lambda}=\nonumber\\
&&\hspace{-15mm}\frac{_{1}F_{2}\left([m+1],\left[1,
\lambda+\frac{1}{2}+m\right],\overline{\mathfrak{z'}}\mathfrak{z}\right)}{\sqrt{_{1}F_{2}\left([m+1],\left[1,
\lambda+\frac{1}{2}+m\right],|\mathfrak{z'}|^2\right)\hspace{1mm}_{1}F_{2}\left([m+1],\left[1,
\lambda+\frac{1}{2}+m\right],|\mathfrak{z}|^2\right)}}\nonumber.\end{eqnarray}
$\diamondsuit$ {\bf{\em{Resolution of unity (or completeness)}}}\\
From equation (40) we see that the state $||\bf{\mathfrak{z}},
m\rangle_{\lambda}$ is a linear combination of all number states
starting with $n = m$. In other words, the first $m$ number states
$n = 0, 1, . . ., m - 1$, are absent from these state . Then, the
unity operator in this space is to be written as
\begin{eqnarray}&&\hspace{-20mm}I_{m}^{\lambda}=\sum_{n=m}^{\infty}{|n,\lambda\rangle\langle n, \lambda|}=\sum_{n=0}^{\infty}{|n+m,\lambda\rangle\langle n+m, \lambda|}.\end{eqnarray}
Evidently, in the right-hand side of the above equation, the
identity operator on the full Hilbert space does not appear, because
of the initial $m$ states of the basis set vectors are omitted. This
leads to the following resolution of unity via bounded, positive
definite and non-oscillating measures, $d\eta _{m,}(|\mathfrak{z}|)
:= K_{m}(|\mathfrak{z}|) \frac{d|\mathfrak{z}|^2}{2}d\varphi$,
\begin{eqnarray}&&\hspace{-25mm}\oint_{\mathbb{C}(\mathfrak{z})}{||\bf{\mathfrak{z}}, m\rangle_{\lambda}
\hspace{1mm}{_{\lambda}\langle \bf{\mathfrak{z}}, m||}}d\eta_{m, \lambda}(|\mathfrak{z}|)=I_{m}^{\lambda},\\
&&\hspace{-25mm}{K}_{m, \lambda}(|\mathfrak{z}|)=
\frac{\Gamma(m+1)}{2\pi\Gamma(\lambda+\frac{1}{2}+m)}{{M}_{m,
\lambda}}(|\mathfrak{z}|)\mathbf{G}^{3 1}_{2
4}\left(|\mathfrak{z}|^2 \mid \hspace{1mm}^{0,\hspace{1mm}
m}_{0,\hspace{1mm} 0,\hspace{1mm} \lambda-\frac{1}{2}+m,\hspace{1mm}
0}\right).\end{eqnarray}
$\diamondsuit$ {\bf{\em{Construction of Nonlinearity function}}}\\
In this section we construct the explicit form of the operator
valued of nonlinearity function associated to the PABGCS. Since, the
coherent states $|{\mathfrak{z}}, \mathfrak{f}_{m}\rangle_{\lambda}$
satisfy following eigenvalue equation
\begin{eqnarray}&&\hspace{-20mm}\left(1+\frac{m}{\hat{N}+\lambda+\frac{1}{2}}\right){J}_{-}^{\lambda}|{\mathfrak{z}}, \mathfrak{f}_{m}\rangle_{\lambda}=\mathfrak{z}|{\mathfrak{z}}, \mathfrak{f}_{m}\rangle_{\lambda},\nonumber\end{eqnarray}
so, multiplying both sides of this equation by
$\left({J}_{+}^{\lambda}\right)^{m}$ yields
\begin{eqnarray}&&\hspace{-20mm}\left({J}_{+}^{\lambda}\right)^{m}\left(1+\frac{m}{\hat{N}+\lambda+\frac{1}{2}}\right){J}_{-}^{\lambda}|{\mathfrak{z}}, \mathfrak{f}_{m}\rangle_{\lambda}=\left({J}_{+}^{\lambda}\right)^{m}\mathfrak{z}|{\mathfrak{z}}, \mathfrak{f}_{m}\rangle_{\lambda}\nonumber\end{eqnarray}
which, making use of the commutation relations (4) and the
identities
\begin{eqnarray}&&\hspace{-20mm}\left({J}_{+}^{\lambda}\right)^{m}{J}_{-}^{\lambda}={J}_{-}^{\lambda}\left({J}_{+}^{\lambda}\right)^{m}
-2m\left(\hat{N}+\frac{\lambda}{2}+\frac{3}{4}-\frac{m}{2}\right)\left({J}_{+}^{\lambda}\right)^{m-1},\nonumber\\
&&\hspace{-20mm}\frac{1}{(\hat{N}+1)(\hat{N}+\lambda+\frac{1}{2})}{J}_{-}^{\lambda}{J}_{+}^{\lambda}=1,\nonumber\end{eqnarray}
leads to
\begin{eqnarray}&&\hspace{-18mm}\frac{\hat{N}+\lambda+\frac{1}{2}}{\hat{N}+\lambda+\frac{1}{2}-m}
\left(1-\frac{2m\left(\hat{N}+\frac{\lambda}{2}+\frac{3}{4}-\frac{m}{2}\right)}
{(\hat{N}+1)(\hat{N}+\lambda+\frac{1}{2})}\right){J}_{-}^{\lambda}||\bf{\mathfrak{z}},
m\rangle_{\lambda}=\mathfrak{z}||\bf{\mathfrak{z}},
m\rangle_{\lambda},\end{eqnarray} and pretends them as nonlinear
coherent states by the expression for nonlinearity function as
\begin{figure}
\begin{center}
\epsfig{figure=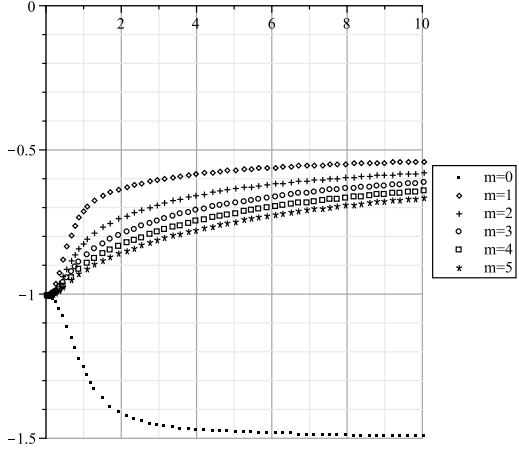,width=7cm}
\end{center}
\caption{\footnotesize Mandel's parameters
${Q}_{m}^{\lambda}(|\mathfrak{z}|)$ in the PABGCSs, versus
$|\mathfrak{z}|$ for different values of $m$ and
$\lambda=\frac{1}{2}$. The dotted curve corresponds to the standard
B-G coherent states.}
\end{figure}
\begin{eqnarray}&&\hspace{-20mm}\frac{\hat{N}+\lambda+\frac{1}{2}}{\hat{N}+\lambda+\frac{1}{2}-m}
\left(1-\frac{2m\left(\hat{N}+\frac{\lambda}{2}+\frac{3}{4}-\frac{m}{2}\right)}
{(\hat{N}+1)(\hat{N}+\lambda+\frac{1}{2})}\right).\end{eqnarray}
Obviously, it transforms to the identity operator for $m=0$. \\\\
$\diamondsuit$ {\bf{\em{Sub-Poissonian Distribution For The Field In PABGCSs }}}\\
In this part we calculate quasi-probablity distribution functions
for the states PABGCS, which provide a convenient way of studying
the nonclassical properties of the field. For this reason we begin
to calculate the expectation values of the number operator $\hat{N}$
and it's square in the basis of the Fock states
\begin{eqnarray}
&&\hspace{-20mm}{\langle{\hat{N}\rangle}_{m}^{\lambda}}=m\hspace{2mm}
\frac{{_{2}F_{3}\left(\left[m+1,m+1\right],\left[1, m, \lambda+\frac{1}{2}+m\right],|\mathfrak{z}|^{2}\right)}}{_{1}F_{2}\left([m+1],\left[1, \lambda+\frac{1}{2}+m\right],|\mathfrak{z}|^2\right)},\nonumber\\
&&\hspace{-20mm}\langle{\hat{N}}^2\rangle_{m}^{\lambda}=m^2\hspace{3mm}
\frac{{_{3}F_{4}\left(\left[m+1, m+1, m+1\right],\left[1, m, m,
\lambda+\frac{1}{2}+m\right],|\mathfrak{z}|^{2}\right)}}{_{1}F_{2}\left([m+1],\left[1,
\lambda+\frac{1}{2}+m\right],|\mathfrak{z}|^2\right)},\nonumber\end{eqnarray}
which result the Mandel parameter $Q_{m}^{\lambda}(|\mathfrak{z}|)$
for these states as follow:
\begin{eqnarray}
&&\hspace{-14mm}Q_{m}^{\lambda}(|\mathfrak{z}|)=
\frac{\langle{\hat{N}}^2\rangle_{m}^{\lambda}-{\langle{\hat{N}\rangle}_{m}^{\lambda}}^{2}
}{{\langle{\hat{N}\rangle}_{m}^{\lambda}}}-1.
\end{eqnarray}
\begin{figure}
\begin{center}
\epsfig{figure=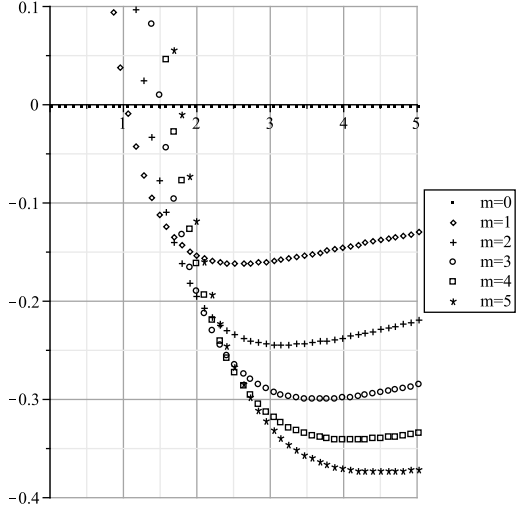,width=7cm}
\end{center}
\caption{\footnotesize Squeezing in $X_{1}^{\lambda}$ quadratures in
the PABGCSs, against $|\mathfrak{z}|$ for $\varphi=0,
\lambda=\frac{1}{2}$ and different values of $m$. The dotted curve
corresponds the standard B-G coherent states( ${S}^{m=0,
\lambda}_{1}=0$).}
\end{figure}
Because of the structure of this function as illustrated in figure 4, the states $||\bf{\mathfrak{z}}, m\rangle_{\lambda}$ show sub-Poissonian statistics( or fully anti-bunching effects).\\\\
$\diamondsuit$ {\bf{\em{Squeezing Properties}}}\\
Our final step is to reveal that measurements on the states
$||\bf{\mathfrak{z}}, m\rangle_{\lambda}$ come with squeezing for
the field quadrature operators $X_{1(2)}^{\lambda}$. In this case,
squeezing factors
\begin{eqnarray}
&&\hspace{-14mm}{S}^{m, \lambda}_{1(2)}=\frac{2\langle(\Delta
X_{1(2)}^{\lambda})^2\rangle}{{|\langle{J}_{3}^{\lambda}
\rangle|}}-1,\nonumber\end{eqnarray} are coming with respect to the
following expectation values
\begin{eqnarray}&&\hspace{-12mm}\left\langle{J}_{+}^{\lambda}\right\rangle=\overline{\left\langle{J}_{-}^{\lambda}\right\rangle}=
\overline{\mathfrak{z}}(m+1)
\frac{{_{1}F_{2}\left(\left[m+2\right],\left[2,\lambda+\frac{1}{2}+m\right],|\mathfrak{z}|^{2}\right)}}
{_{1}F_{2}\left(\left[m+1\right],\left[1, \lambda+\frac{1}{2}+m\right],|\mathfrak{z}|^2\right)},\nonumber\\
&&\hspace{-12mm}\left\langle{{J}_{+}^{\lambda}}^{2}\right\rangle=\overline{\left\langle{{J}_{-}^{\lambda}}^{2}\right\rangle}=
{\overline{\mathfrak{z}}}^{2}\frac{(m+1)(m+2)}{2}
\frac{{_{1}F_{2}\left(\left[m+3\right],\left[3, \lambda+\frac{1}{2}+m\right],|\mathfrak{z}|^{2}\right)}}{_{1}F_{2}\left(\left[m+1\right],\left[1, \lambda+\frac{1}{2}+m\right],|\mathfrak{z}|^2\right)},\nonumber\\
&&\hspace{-12mm}\left\langle{{J}_{+}^{\lambda}}{J}_{-}^{\lambda}\right\rangle=
m\left(m+\lambda-\frac{1}{2}\right)\frac{{_{2}F_{4}\left(\left[m+1,m+1\right],\left[1, m, \lambda-\frac{1}{2}+m\right],|\mathfrak{z}|^{2}\right)}}{_{1}F_{2}\left(\left[m+1\right],\left[1, \lambda+\frac{1}{2}+m\right],|\mathfrak{z}|^2\right)},\nonumber\\
&&\hspace{-12mm}\left\langle{J}_{3}^{\lambda}\right\rangle=m\hspace{2mm}
\frac{{_{2}F_{3}\left(\left[m+1,m+1\right],\left[1, m,
\lambda+\frac{1}{2}+m\right],|\mathfrak{z}|^{2}\right)}}{_{1}F_{2}\left([m+1],\left[1,
\lambda+\frac{1}{2}+m\right],|\mathfrak{z}|^2\right)}+\frac{\lambda}{2}+\frac{1}{4}.\nonumber\end{eqnarray}
Figure 5 implies that the squeezing in $X_{1}$ arises when  $m$ is
increasing, but it is disappeared while $m\rightarrow 0$.
\section{Conclusions}
Based on the process proposed in this presentation, it will be
possible to reproduce broad range of states that are called
nonlinear coherent states through of the generalized displacement
operators. In a general view the formalism presented here provides a
unified approach to construct all the employed CSs already
introduced in different ways( the Barut-Girardello and  nonlinear
coherent states). These states realize a resolution of the identity
with respect to positive definite measures on the complex plane.
Finally, non-classicality properties of such states have been
reviewed in detail. For instance, we have shown that their squeezing
in $X_{1}^{\lambda}$ quadratures are really considerable. In other
words the system can be prepared in any excited Fock states so that
${\mathcal{S}}^{m\geq 1, \lambda}_{1(2)}< 0$.\\{\em{As significant pictures of the above mentioned approach:\\
$\bullet$ We model quantum mechanical system which follows fully sub-poissonian statistics.\\
$\bullet$ Performance of a strategy for finding considerable as well as controllable squeezing properties in field quadratures.\\
$\bullet$ Non-classical features of these states could be strictly
controlled by energy number.}}\\ We assert that our approach, has
the potentiality to be used for the construction of a variety of new
classes of generalized nonlinear coherent states, corresponding to
some, both physical and mathematical, solvable quantum system, such
as half oscillator, radial part of a 3D harmonic oscillator and so
on, with known discrete and (non-) degenerate spectrum. Especially,
one can suggest a particularization: $\lambda=2k-\frac{1}{2}$, where
$k$ is the Bargmann index labeling the IREP. Inserting this value in
Eq. (1), we obtain the radial part of the Hamiltonian of the
pseudo-harmonic oscillator. Hence, a series of results obtained in
this document can be found, as particular case, in \cite{Popov,
Popov1, Dong}.

An example of application of this technique, together with
procedures provided in \cite{Sander}, could include design entangled
nonlinear Barut- Giradello coherent states( ENBGCSs) as
superpositions of multi-particle NBGCSs. ENBGCSs are coherent states
with respect to generalized $su(1,1)$ generators, and multi-particle
parity states arise as a special case. Also, based on our
calculations, the entangled negative binomial states and entangled
squeezed states can be considered as kind of ENBGCSs. Quantum
discord as well as degrees of entanglement are calculated, which are
obviously depends on quantum number $m$ and complex variable $z$.\\
$\bullet$ {\em Finally, we conclude that maximally entanglement will
be achieved when the squeezing in both of field quadratures tend to
zero. In other word, squeezing may be a good candidate as one of
maximally entanglement witnesses.}\\\\ \textbf{Acknowledgments}\\
The author would like to thank the referees for their interesting
and worthwhile suggestions to improve the presentation.

\end{document}